\PassOptionsToPackage{unicode}{hyperref}
\PassOptionsToPackage{hyphens}{url}
\PassOptionsToPackage{dvipsnames,svgnames,x11names}{xcolor}
\documentclass[
  letterpaper,
  DIV=11,
  numbers=noendperiod]{scrartcl}
\usepackage{xcolor}
\usepackage{amsmath}
\usepackage[T1]{fontenc}
\usepackage{stix2}
\IfFileExists{upquote.sty}{\usepackage{upquote}}{}
\IfFileExists{microtype.sty}{
  \usepackage[]{microtype}
  \UseMicrotypeSet[protrusion]{basicmath} 
}{}
\usepackage{setspace}
\makeatletter
\@ifundefined{KOMAClassName}{
  \IfFileExists{parskip.sty}{%
    \usepackage{parskip}
  }{
    \setlength{\parindent}{0pt}
    \setlength{\parskip}{6pt plus 2pt minus 1pt}}
}{
  \KOMAoptions{parskip=half}}
\makeatother
\makeatletter
\ifx\paragraph\undefined\else
  \let\oldparagraph\paragraph
  \renewcommand{\paragraph}{
    \@ifstar
      \xxxParagraphStar
      \xxxParagraphNoStar
  }
  \newcommand{\xxxParagraphStar}[1]{\oldparagraph*{#1}\mbox{}}
  \newcommand{\xxxParagraphNoStar}[1]{\oldparagraph{#1}\mbox{}}
\fi
\ifx\subparagraph\undefined\else
  \let\oldsubparagraph\subparagraph
  \renewcommand{\subparagraph}{
    \@ifstar
      \xxxSubParagraphStar
      \xxxSubParagraphNoStar
  }
  \newcommand{\xxxSubParagraphStar}[1]{\oldsubparagraph*{#1}\mbox{}}
  \newcommand{\xxxSubParagraphNoStar}[1]{\oldsubparagraph{#1}\mbox{}}
\fi
\makeatother

\usepackage{longtable,booktabs,array}
\usepackage{calc} 
\usepackage{etoolbox}
\makeatletter
\patchcmd\longtable{\par}{\if@noskipsec\mbox{}\fi\par}{}{}
\makeatother
\IfFileExists{footnotehyper.sty}{\usepackage{footnotehyper}}{\usepackage{footnote}}
\makesavenoteenv{longtable}
\usepackage{graphicx}
\makeatletter
\newsavebox\pandoc@box
\newcommand*\pandocbounded[1]{
  \sbox\pandoc@box{#1}%
  \Gscale@div\@tempa{\textheight}{\dimexpr\ht\pandoc@box+\dp\pandoc@box\relax}%
  \Gscale@div\@tempb{\linewidth}{\wd\pandoc@box}%
  \ifdim\@tempb\p@<\@tempa\p@\let\@tempa\@tempb\fi
  \ifdim\@tempa\p@<\p@\scalebox{\@tempa}{\usebox\pandoc@box}%
  \else\usebox{\pandoc@box}%
  \fi%
}
\def\fps@figure{htbp}
\makeatother

\NewDocumentCommand\citeproctext{}{}

\makeatletter
 \let\@cite@ofmt\@firstofone
 \def\@biblabel#1{}
 \def\@cite#1#2{{#1\if@tempswa , #2\fi}}
\makeatother
\newlength{\cslhangindent}
\newlength{\csllabelwidth}
\newenvironment{CSLReferences}[2] 
 {\begin{list}{}{%
  \setlength{\itemindent}{0pt}
  \setlength{\leftmargin}{0pt}
  \setlength{\parsep}{0pt}
  \ifodd #1
   \setlength{\leftmargin}{\cslhangindent}
   \setlength{\itemindent}{-1\cslhangindent}
  \fi
  \setlength{\itemsep}{#2\baselineskip}}}
 {\end{list}}
\usepackage{calc}

\newcommand{\CSLLeftMargin}[1]{\parbox[t]{\csllabelwidth}{\strut#1\strut}}
\newcommand{\CSLRightInline}[1]{\parbox[t]{\linewidth - \csllabelwidth}{\strut#1\strut}}

\KOMAoption{captions}{tableheading}
\makeatletter
\@ifpackageloaded{caption}{}{\usepackage{caption}}
\AtBeginDocument{%
\ifdefined\contentsname
  \renewcommand*\contentsname{Table of contents}
\else
  \newcommand\contentsname{Table of contents}
\fi
\ifdefined\listfigurename
  \renewcommand*\listfigurename{List of Figures}
\else
  \newcommand\listfigurename{List of Figures}
\fi
\ifdefined\listtablename
  \renewcommand*\listtablename{List of Tables}
\else
  \newcommand\listtablename{List of Tables}
\fi
\ifdefined\figurename
  \renewcommand*\figurename{Figure}
\else
  \newcommand\figurename{Figure}
\fi
\ifdefined\tablename
  \renewcommand*\tablename{Table}
\else
  \newcommand\tablename{Table}
\fi
}
\@ifpackageloaded{float}{}{\usepackage{float}}
\floatstyle{ruled}
\@ifundefined{c@chapter}{\newfloat{codelisting}{h}{lop}}{\newfloat{codelisting}{h}{lop}[chapter]}
\floatname{codelisting}{Listing}

\makeatother
\makeatletter
\@ifpackageloaded{caption}{}{\usepackage{caption}}
\@ifpackageloaded{subcaption}{}{\usepackage{subcaption}}
\makeatother
\usepackage{bookmark}
\IfFileExists{xurl.sty}{\usepackage{xurl}}{} 
\hypersetup{
  pdftitle={Predicting transcriptional interactions with GrnPFN},
  colorlinks=true,
  linkcolor={blue},
  filecolor={Maroon},
  citecolor={Blue},
  urlcolor={Blue},
  pdfcreator={LaTeX via pandoc}}

\title{Predicting transcriptional interactions with GrnPFN}
\author{}
\date{}
\begin{document}
\maketitle

\setstretch{1.15}
Haihui Zhang\textsuperscript{1,2*}

\textsuperscript{1} Department of Orofacial Sciences, School of
Dentistry, University of California San Francisco, San Francisco,
California, United States of America

\textsuperscript{2} Cancer Biology and Immunology Laboratory, University
of California San Francisco, San Francisco, California, United States of
America

* Corresponding author

E-mail: Haihui.Zhang@ucsf.edu (HZ)

Short title: Predicting interactions in sparse perturbation screens

\newpage{}

\section{Abstract}\label{abstract}

Combinatorial therapies offer opportunities to improve treatment
efficacy and overcome therapeutic resistance, but their rational design
requires understanding how combined interventions interact beyond their
individual effects. Combined genetic and chemical perturbations can
elicit nonadditive transcriptional responses, yet the vast number of
possible combinations limits experimental characterization. Predicting
these interactions is particularly challenging when few or no
combinations have been measured. Here we present GrnPFN, a prior-data
fitted network that predicts nonadditive transcriptional responses to
combinations from single-perturbation profiles. Pretrained on simulated
screens generated from random gene-regulatory networks, GrnPFN
incorporates measured combinations through in-context learning without
retraining. We evaluated GrnPFN in single-cell screens spanning CRISPR
activation, RNA knockdown and drug combinations, using separate cells to
construct model inputs, training targets and test targets. Without
measured combinations in the target screen, GrnPFN outperformed gradient
boosting trained on other screens. With limited combination
measurements, it outperformed kernel ridge regression fitted to the same
measured pairs. These advantages extended to an additional screen
combining CRISPR interference with a drug. Our results demonstrate that
pretraining on simulated regulatory networks enables prediction of
nonadditive transcriptional interactions across perturbation modalities
when experimental measurements are scarce.

\section{Introduction}\label{introduction}

Cellular responses to combined perturbations can depart from the sum of
their individual effects, revealing interactions that are not apparent
from single perturbations alone. Single-cell genetic and chemical
screens measure these responses at transcriptome scale {[}1--3{]}, and
combinatorial studies have characterized interactions involving genes,
RNA perturbations and drugs {[}4--7{]}. Yet the number of possible pairs
grows quadratically with the number of perturbations, leaving most
combinations unmeasured. The practical challenge is therefore to predict
interactions from single-perturbation profiles before combination
measurements are available, and to incorporate the limited measurements
obtained as a screen progresses.

Methods including GEARS, the compositional perturbation autoencoder
(CPA) and CellFlow predict transcriptional responses to unmeasured
perturbations or combinations {[}8--10{]}. Assessing their contribution
requires distinguishing prediction of the full expression response from
prediction of the interaction itself. Recent benchmarks have shown that
simple baselines can be competitive with deep-learning models {[}11{]},
while calibration-aware analyses have identified settings in which more
complex models provide additional predictive value {[}12{]}. In the K562
combinatorial screen, adding single-perturbation effects captures much
of the predictive accuracy obtained for full expression responses.
Predicting the residual beyond this additive expectation directly
targets the nonadditive component. However, regression models fitted to
measured pairs have limited evidence from which to learn this component
when combinations are sparsely sampled.

Informed modeling incorporates biological information in a form suited
to the inference task. For example, scBIOT can use cell-identity
annotations to guide optimal-transport alignment across single-cell
datasets {[}13{]}. For perturbation prediction, biological assumptions
can be encoded in a distribution of simulated datasets used for
pretraining. Prior-data fitted networks (PFNs) learn predictive
structure from datasets sampled from such a prior and subsequently
condition on observations from a new dataset through in-context learning
{[}14--16{]}. This framework allows a model to draw on relationships
learned before application to the target dataset and to incorporate new
observations without retraining.

Predicting nonadditive transcriptional responses to combined
perturbations is essential for understanding how molecular interventions
interact and for guiding the rational design of combinatorial therapies.
Recent applications include MapPFN, which learns from simulated
gene-regulatory networks (GRNs) to predict genetic perturbation
responses {[}17{]}, and PerturbPFN, which uses synthetic structural
models for chemical perturbation prediction {[}18{]}. General-purpose
tabular foundation models have also been evaluated on
single-perturbation responses {[}19{]}. Whereas these approaches focus
on responses to individual perturbations, GrnPFN directly predicts the
nonadditive effects of combinations from single-perturbation profiles,
including when no combinations have been measured in the target screen.
For combinatorial prediction, the prior can be designed specifically to
relate measured single-perturbation effects to departures from
additivity. GRN simulations can represent how regulatory saturation and
convergence on shared targets generate these departures, providing
interaction examples beyond those measured experimentally.

Here we introduce GrnPFN, a GRN-informed PFN pretrained specifically to
predict nonadditive transcriptional responses from measured
single-perturbation profiles. Pretraining on genetic and chemical
combinations simulated from random GRNs provides an interaction prior
before any pairs in the target screen have been measured. The same
pretrained model makes predictions without measured pairs and
incorporates available pairs through in-context learning without
updating its weights. This supplies predictive structure that regression
models fitted only to target-screen pairs must estimate from limited
experimental observations. GrnPFN uses no gene-identity embeddings and
requires no reconstructed regulatory network for the target cell type,
allowing the same model to be applied across the genetic and chemical
screens evaluated here.

In pooled analyses across five single-cell screens, GrnPFN outperformed
gradient boosting trained on other screens when no target-screen
combinations were available, and kernel ridge regression fitted to the
same measured pairs when combination data were limited. The advantage
extended to test pairs involving perturbations absent from the
measured-pair context. With all training pairs available in K562, GrnPFN
also achieved higher interaction-prediction accuracy than GEARS, CPA and
CellFlow. Its advantage at low measurement budgets transferred without
retraining to an additional CRISPR interference--drug screen, a
perturbation design absent from pretraining. These results demonstrate
that a model trained on simulated regulatory dynamics can improve
transcriptional interaction prediction at the beginning of combinatorial
screens, when empirical combination data are most limited.

\section{Results}\label{results}

\subsection{GrnPFN learns a transferable interaction prior from
simulated
screens}\label{grnpfn-learns-a-transferable-interaction-prior-from-simulated-screens}

We defined the effect of perturbation \(p\), \(\Delta_p\), as its mean
change in log-normalized gene expression relative to control cells. For
a pair of perturbations A and B, the interaction was

\[R_{AB}=\Delta_{AB}-\Delta_A-\Delta_B.\]

This residual describes the component of the combination response that
is not explained by adding the two single effects (Fig 1a). GrnPFN
predicts the residual directly, allowing its contribution to be
evaluated separately from the additive response.

\includegraphics[width=1\linewidth,height=\textheight,keepaspectratio]{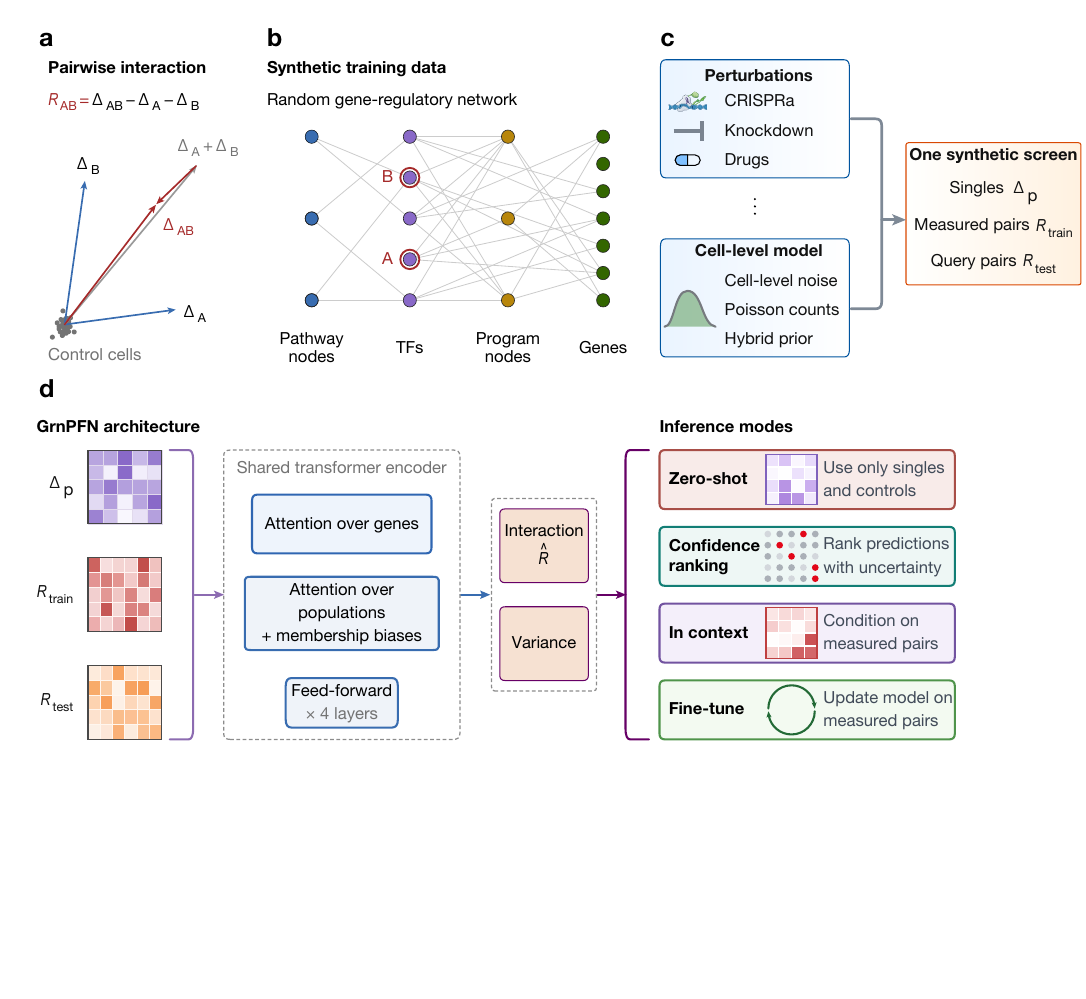}

\textbf{Fig 1. Simulation-pretrained prediction of transcriptional
interactions.} a) The interaction residual \(R_{AB}\) is the difference
between the pair response \(\Delta_{AB}\) and the sum of the
single-perturbation responses \(\Delta_A\) and \(\Delta_B\), each
measured relative to control cells. b) Pretraining used 140,000
simulated screens generated from random gene-regulatory networks
containing pathway nodes, transcription factors (TFs), program nodes,
and measured genes. A and B mark a perturbed pair of transcription
factors. c) Each simulated screen combines CRISPR activation, knockdown,
or drug perturbations with cell-level noise, Poisson count sampling, and
a hybrid prior that adds a screen-specific interaction component. A
screen provides single-perturbation effects \(\Delta_p\),
training-target interactions of measured pairs \(R_{\mathrm{train}}\),
and test-target interactions of query pairs \(R_{\mathrm{test}}\). d)
GrnPFN arranges single perturbations, measured pairs, and query pairs as
a grid of populations by genes. Query pairs carry only features of their
members' single effects, while their interactions serve as prediction
targets. A shared transformer encoder applies attention over genes,
attention over populations with membership biases, and a feed-forward
block in each of four layers. Query populations do not serve as
attention keys. Separate heads predict interactions \(\hat{R}\) and
predictive variances. The model supports zero-shot prediction from
single-perturbation and control profiles, confidence ranking by relative
uncertainty, in-context prediction conditioned on measured pairs, and
fine-tuning on measured pairs.

Pretraining used simulated screens generated from random gene-regulatory
networks comprising unmeasured pathway nodes, perturbable transcription
factors, saturating program nodes, and measured genes (Fig 1b). The
simulator included CRISPR activation, knockdown, and drug combinations,
with variability in cell state, perturbation strength, and count
sampling. A hybrid prior supplemented the network-derived response with
a screen-specific interaction component (Fig 1c). Because this component
varied between screens, measured pairs provided information that could
not be recovered from the single profiles alone.

GrnPFN represents a screen as a grid of populations by genes. Four
layers alternate attention across genes and populations, with learned
attention biases linking pairs to their members and to other pairs
sharing a member (Fig 1d). Measured interactions enter as context, and
unmeasured pairs receive predictions from an interaction head and a
predictive-variance head. The same pretrained model supports zero-shot
prediction from single-perturbation profiles and in-context prediction
using measured pairs. A copy of the model can also be fine-tuned on
measured pairs (Fig 1d). The main prediction benchmark used frozen model
weights trained entirely on simulated data.

\subsection{Separating cell roles enables evaluation of interaction
prediction}\label{separating-cell-roles-enables-evaluation-of-interaction-prediction}

Interaction targets contain the negative of each single-perturbation
effect. Consequently, when the same cells provide a single-profile input
and its contribution to the interaction target, sampling error enters
the input and target with opposite signs. This coupling can produce high
apparent accuracy without recovery of the underlying interaction. In a
melanoma Perturb-CITE-seq dataset {[}20{]}, a predictor that assumed the
knockout effect disappeared under interferon-\ensuremath{\gamma} achieved a median cosine
similarity of 0.899 when inputs and targets shared cells, but \ensuremath{-}0.006
when they were calculated from disjoint cells.

We therefore assigned single-perturbation and control cells to separate
thirds for model inputs, training targets, and test targets (Methods).
Pairs were assigned to five folds, with each pair held out once. The
cells belonging to a pair were used together because that pair served
either training or testing within a fold. This design separated the
single-perturbation and control-cell noise across the three roles while
retaining the available depth for each combination.

The main benchmark included CRISPR activation in K562 cells, Cas13
knockdown in THP-1 cells, and drug combinations in A172, T98G, and U87MG
cells (Table 1). Simulator and architecture development used one cell
partition of the first four screens, and U87MG was excluded from these
choices. We then evaluated the frozen pipeline on four new partitions of
the existing cells. Predictions were scored by cosine similarity over
the 100 genes with the largest absolute observed interaction after
centering. Nonzero-budget benchmark comparisons used measured training
pairs to center observed interactions. Learning curves used test-pair
centering throughout, which kept the observed target constant across
measurement budgets (Methods).

\textbf{Table 1. Perturbation screens used to evaluate GrnPFN.}

{\def\LTcaptype{none} 
\begin{longtable}[]{@{}
  >{\raggedright\arraybackslash}p{(\linewidth - 6\tabcolsep) * \real{0.2500}}
  >{\raggedright\arraybackslash}p{(\linewidth - 6\tabcolsep) * \real{0.2500}}
  >{\raggedleft\arraybackslash}p{(\linewidth - 6\tabcolsep) * \real{0.2500}}
  >{\raggedright\arraybackslash}p{(\linewidth - 6\tabcolsep) * \real{0.2500}}@{}}
\toprule\noalign{}
\begin{minipage}[b]{\linewidth}\raggedright
Screen
\end{minipage} & \begin{minipage}[b]{\linewidth}\raggedright
Perturbation design
\end{minipage} & \begin{minipage}[b]{\linewidth}\raggedleft
Eligible pairs
\end{minipage} & \begin{minipage}[b]{\linewidth}\raggedright
Evaluation role
\end{minipage} \\
\midrule\noalign{}
\endhead
\bottomrule\noalign{}
\endlastfoot
K562 & CRISPR activation pairs & 131 & Model development and main
benchmark \\
THP-1 & Cas13 knockdown pairs & 142 & Model development and main
benchmark \\
A172 & Drug combinations with trametinib & 82 & Model development and
main benchmark \\
T98G & Drug combinations with trametinib & 84 & Model development and
main benchmark \\
U87MG & Drug combinations with trametinib & 81 & Cell line excluded from
model development \\
A375 & CRISPR interference with vemurafenib & 142 & Additional
drug--gene evaluation \\
\end{longtable}
}

Counts denote pairs meeting the cell-count threshold for the pair and
both single conditions. Single-perturbation profiles were available for
every pair member.

\subsection{Simulation pretraining improves prediction at low
measurement
budgets}\label{simulation-pretraining-improves-prediction-at-low-measurement-budgets}

Without measured pairs in the target screen, GrnPFN outperformed
gradient boosting trained on the other four screens, with a pooled
cosine similarity difference of 0.049 (97.5\% CI, 0.038--0.060). The
largest gains occurred in K562, A172, and U87MG, while performance was
similar in THP-1 and T98G (Fig 2a,b). GrnPFN also outperformed a
transferred linear subadditivity model.

\includegraphics[width=1\linewidth,height=\textheight,keepaspectratio]{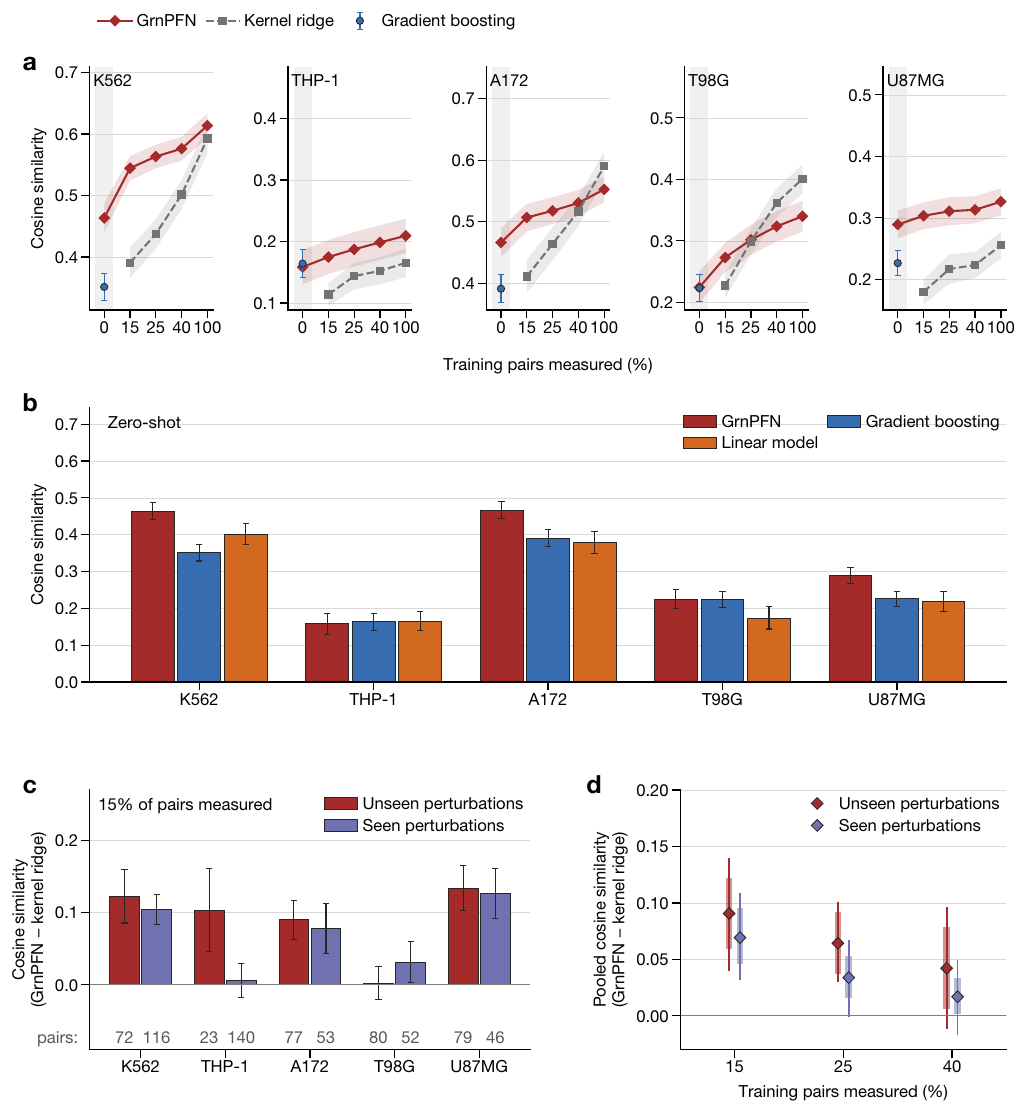}

\textbf{Fig 2. GrnPFN improves interaction prediction at low measurement
budgets.} a) Learning curves for GrnPFN and kernel ridge, with
transferred gradient boosting at zero measured pairs. Partial budgets
use three random draws of 15\%, 25\%, or 40\% of each fold's training
pairs, with a minimum of 10 measured pairs. The 100\% budget includes
all training pairs. Test-pair centering gives each held-out pair the
same observed target across budgets. Budget positions are equally
spaced, and panels have equal vertical spans with different limits. b)
Zero-shot comparisons with gradient boosting and a linear subadditivity
model trained on the other four screens. c) GrnPFN minus kernel ridge at
15\%, grouped by perturbation coverage and scored with benchmark
centering. Unseen genetic pairs have neither member in measured context,
whereas unseen drug pairs have their non-anchor compound absent at every
dose. Numbers indicate unique contributing pairs across draws and cell
partitions, and groups can overlap. d) Equal-screen pooled differences
by budget, including screens with at least five pairs in a group. THP-1
is excluded from the unseen group at 25\% and 40\%. Diamonds denote
means. Thick and thin bars denote pair-bootstrap and dependence-aware
95\% intervals, respectively. Panels a--c show means and standard errors
over pairs after averaging applicable draws and the four cell
partitions.

When 15\% of training pairs were measured, GrnPFN outperformed kernel
ridge regression fitted to the same pairs, with a pooled difference of
0.069 (95\% CI, 0.048--0.090) using benchmark centering (Methods). With
the test-pair centering used for the learning curves, the corresponding
difference was 0.096 (95\% CI, 0.080--0.112) (Fig 2a). This advantage
decreased as more pairs became available, and the two methods achieved
comparable overall performance when all training pairs were measured
(Fig 2a). These results indicate that the simulation-trained prior is
most valuable when few or no combination measurements are available.

\subsection{GrnPFN generalizes beyond perturbations represented in
measured
combinations}\label{grnpfn-generalizes-beyond-perturbations-represented-in-measured-combinations}

Sparse measurement budgets leave some perturbations absent from the
measured combinations. We classified a genetic test pair as unseen when
neither member occurred in a measured pair, and a drug pair as unseen
when its non-anchor compound was absent at every dose.
Single-perturbation profiles were available in both cases. This analysis
tested whether the model could predict beyond the perturbations
represented in its combination context.

At the 15\% budget (Fig 2c), GrnPFN exceeded kernel ridge by 0.091 for
unseen pairs (95\% CI, 0.059--0.122) and by 0.069 for seen pairs
(0.046--0.095). Dependence-aware intervals remained positive for both
groups (0.040--0.139 and 0.033--0.108). In THP-1, the differences were
0.103 for unseen pairs and 0.006 for seen pairs. The corresponding
pooled differences at 25\% were 0.064 and 0.034, and at 40\% were 0.042
and 0.017 (Fig 2d). At 40\%, dependence-aware intervals included zero.
These results indicate that the early-screen advantage extended to pairs
whose members had not yet been measured in combination. Comparisons
between the groups were descriptive because their membership changed
across measurement draws.

\subsection{GrnPFN recovers strong interaction patterns in K562
cells}\label{grnpfn-recovers-strong-interaction-patterns-in-k562-cells}

To examine individual predictions, we inspected the K562 pairs with the
largest observed interactions. The eight strongest pairs involving
\emph{CEBPA}, \emph{CEBPB}, or \emph{CEBPE} had zero-shot cosine
similarities of 0.65--0.94 (Fig 3a). The most accurately predicted of
these, \emph{CEBPA} + \emph{CEBPE}, reached a cosine similarity of 0.94,
with agreement in the direction of gene-level deviations from additivity
(Fig 3b).

\includegraphics[width=1\linewidth,height=\textheight,keepaspectratio]{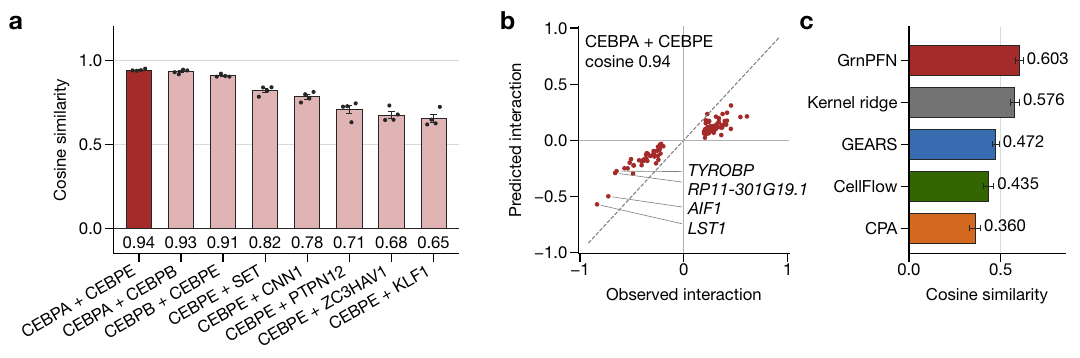}

\textbf{Fig 3. Interaction patterns and full-budget prediction in K562
cells.} a) Zero-shot predictions for the eight pairs involving
\emph{CEBPA}, \emph{CEBPB}, or \emph{CEBPE} with the largest observed
interactions, ordered by cosine similarity. Bars show means across four
cell partitions, points show individual partitions, and error bars show
their standard errors. b) Observed and predicted interactions for
\emph{CEBPA} + \emph{CEBPE}, averaged across partitions over the 100
selected genes. The dashed line indicates identity, and labels identify
the four largest observed deviations. Panels a and b use test-pair
centering. c) Prediction with all training pairs, using benchmark
centering. Bars show means over pairs and error bars their standard
errors. Deep-learning comparators used the same pair folds but all
single-perturbation and control cells (Methods).

We also compared predictions when all training pairs were available. In
K562, mean cosine similarity was 0.603 for GrnPFN, 0.576 for kernel
ridge, 0.472 for GEARS, 0.435 for CellFlow, and 0.360 for CPA (Fig 3c).
Across the other four screens, GrnPFN exceeded CellFlow and CPA by
pooled differences of 0.078 (95\% CI, 0.065--0.092) and 0.111
(0.095--0.126). Its difference from GEARS in THP-1 was 0.044
(0.014--0.073), evaluated on the 127 pairs for which GEARS returned
predictions. Kernel ridge achieved similar advantages over these
deep-learning comparators. All methods used the same held-out pair
folds, although the deep-learning comparators had access to all
single-perturbation and control cells rather than the input third used
by GrnPFN and the regression baselines (Methods).

\subsection{Predictive uncertainty identifies more accurate subsets,
while adaptation varies across
screens}\label{predictive-uncertainty-identifies-more-accurate-subsets-while-adaptation-varies-across-screens}

We ranked zero-shot predictions by relative uncertainty, defined as the
total predicted variance divided by the squared magnitude of the
predicted interaction. In every screen, the most confident third had
higher cosine similarity than the least confident third (Fig 4a), with a
pooled difference of 0.230 (95\% CI, 0.188--0.272). Retaining a smaller
fraction of the most confident pairs generally increased mean accuracy
(Fig 4b).

\includegraphics[width=1\linewidth,height=\textheight,keepaspectratio]{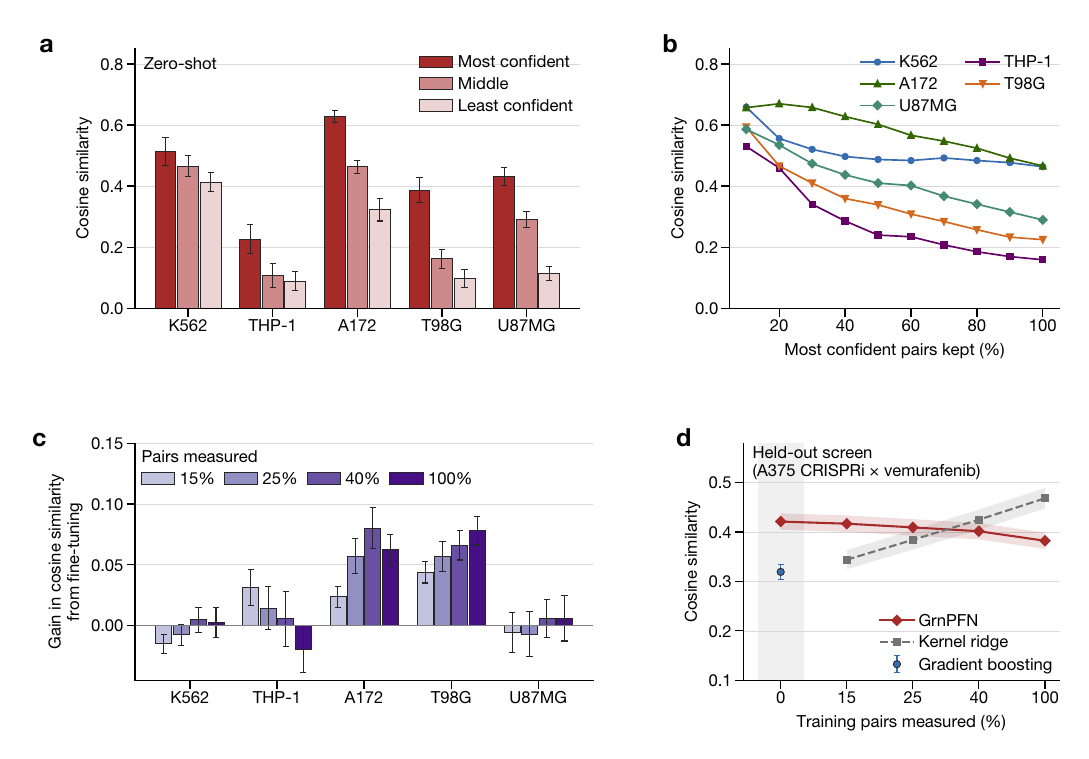}

\textbf{Fig 4. Confidence ranking, fine-tuning, and transfer to the
drug--gene screen.} a) Zero-shot cosine similarity in the most
confident, middle, and least confident thirds, defined within each
screen and cell partition by relative uncertainty, calculated as the
total predicted variance divided by the squared magnitude of the
centered prediction. b) Mean cosine similarity after retaining an
increasing fraction of the most confident pairs. Panels a and b use
test-pair centering and four cell partitions. c) Change in cosine
similarity after fine-tuning, by measurement budget, using benchmark
centering on the development partition. d) Prediction in the additional
A375 CRISPR interference--vemurafenib screen with 142 eligible pairs.
GrnPFN and kernel ridge are evaluated across budgets, and gradient
boosting trained on the five main screens is shown at zero. Test-pair
centering holds the observed target constant across budgets. Budget
positions are equally spaced, with the zero-budget region shaded. Bars
and points denote means over test pairs after applicable averaging.
Error bars and bands denote standard errors over pairs.

Measured context contributed to prediction accuracy. On the development
partition, permuting interaction values among context pairs at the full
measurement budget reduced cosine similarity by 0.226 in K562 and by
0.067 to 0.157 in the other screens. Updating the model weights provided
further gains in some settings. Fine-tuning improved cosine similarity
across budgets by 0.024--0.080 in A172 and 0.044--0.078 in T98G, but had
inconsistent effects in K562, THP-1, and U87MG (Fig 4c). These
development experiments supported retaining frozen in-context prediction
as the main evaluation setting and treating fine-tuning as an optional,
screen-dependent extension.

\subsection{An additional drug--gene screen supports transfer at low
measurement
budgets}\label{an-additional-druggene-screen-supports-transfer-at-low-measurement-budgets}

We next evaluated a perturbation design not represented in the
simulator, combining a genetic perturbation with a drug. Candidate
datasets were required to contain all four conditions needed to estimate
an interaction and to meet a split-half reproducibility threshold for
training pairs before model evaluation. An A375 melanoma screen
combining CRISPR interference (CRISPRi) with vemurafenib passed this
criterion, with a split-half cosine similarity of 0.255 and 142 eligible
pairs {[}21{]}.

GrnPFN generalized to the held-out A375 CRISPRi plus vemurafenib screen,
despite being pretrained on simulations that did not include genetic and
drug combinations. Without any measured pairs or retraining, GrnPFN
achieved a zero-shot cosine similarity of 0.421, outperforming gradient
boosting trained on the five main screens (0.319) by 0.102 (95\% CI,
0.090--0.113) (Fig 4d). With only 15\% of training pairs measured,
GrnPFN also outperformed kernel ridge regression fitted to the same
pairs by 0.044 (95\% CI, 0.021--0.067) using benchmark centering. With
the test-pair centering used in Fig 4d, the corresponding difference was
0.073 (95\% CI, 0.052--0.095). The difference narrowed as more pairs
became available, with kernel ridge achieving higher accuracy at larger
measurement budgets. These results demonstrate that GrnPFN can transfer
to a perturbation design not represented during pretraining and provide
an advantage when combination measurements are limited.

\section{Discussion}\label{discussion}

GrnPFN demonstrates that models pretrained on simulated GRN screens can
generalize to real combinatorial perturbation experiments and predict
nonadditive transcriptional responses. By leveraging a GRN-informed
prior, GrnPFN enables zero-shot prediction from single-perturbation
profiles without requiring measured combinations in the target screen.
In this setting, GrnPFN outperformed models trained on other
experimental screens and, when limited combination measurements were
available, regression models fitted to the same data. This advantage
extended to an independent drug--gene perturbation screen, supporting
the transferability of the simulation-derived prior across perturbation
modalities. However, as more combination measurements became available,
screen-specific regression models generally improved and are comparable
to GrnPFN. These findings highlight the value of GRN-informed
pretraining for predicting combinatorial responses when experimental
measurements are scarce, while indicating that its relative advantage
diminishes as screen-specific data accumulate.

Directly predicting the interaction residual provides a biologically
meaningful measure of nonadditive transcriptional responses. By
isolating deviations from the additive effects of individual
perturbations on the chosen expression scale, this approach evaluates
the ability to predict interactions rather than the larger additive
component. However, transcriptional nonadditivity does not necessarily
indicate a specific molecular mechanism or imply synergistic effects on
other phenotypes, such as cell fitness. Accurate evaluation also
requires careful consideration of measurement noise, as reusing the same
single-perturbation cells to construct model inputs and evaluation
targets can inflate predictive performance through shared sampling
errors. To minimize this bias, we used independent cells for model
inputs, training targets, and test targets, together with comparisons
against established regression baselines and fixed evaluation targets.
This design enables a more rigorous assessment of the predictive
information captured by GrnPFN.

The contrast between zero-shot performance and context adaptation points
to both the utility and the limitations of the simulation prior. The
hybrid interaction component and membership-aware attention increased
context use during development, but did not make additional measurements
uniformly beneficial. In particular, the A375 drug--gene screen
supported zero-shot transfer while showing declining accuracy as context
increased. Fine-tuning recovered additional performance in two drug
screens but did not help consistently across datasets. These results
suggest that transferable relationships between single effects and
interactions can be easier to capture than the screen-specific structure
needed for effective adaptation.

The breadth of the present evaluation is limited by the available
datasets. Model development used one partition of four real screens, and
the four later partitions reused the same cells. U87MG was excluded from
model selection but shared its chemical panel and anchor design with two
development screens. The additional drug--gene evaluation involved one
drug in one cell line. Pair folds also retained substantial overlap in
members or compounds, so they primarily assessed unmeasured combinations
of perturbations with known single effects. Dependence-aware intervals
address some of this structure but do not replace evaluation in
independent biological experiments.

Together, these findings position simulation-trained in-context learning
as a complement to experimental measurement in sparse combinatorial
screens. The model supplies an initial interaction prior that can be
used immediately and revised with measured combinations. Its practical
value lies in reducing the dependence on a large initial combination
training set, while leaving room for fitted models to become preferable
as a screen grows. Prospective evaluation across additional biological
contexts will determine how reliably this early predictive advantage
holds.

\section{Materials and methods}\label{materials-and-methods}

\subsection{Datasets and
preprocessing}\label{datasets-and-preprocessing}

The K562 CRISPR activation screen was obtained from GSE133344 {[}4{]}.
Cells carrying two negative-control guides defined the control
population, one targeting guide defined a single perturbation, and two
targeting guides defined an alphabetically ordered pair. We retained 131
pairs with at least 20 cells in the pair population and in each
corresponding single-perturbation population. Counts were normalized to
10,000 per cell and log-transformed. We selected 3,000 highly variable
genes using the Seurat v3 procedure on single-perturbation and control
cells {[}22,23{]}.

The THP-1 Cas13 screen was obtained from GSE213957 {[}6{]}.
Perturbations were assigned from guide pairs, and preprocessing followed
the K562 procedure, with highly variable genes selected using a random
subset of 20,000 cells. The same cell-count threshold retained 142
pairs. Drug screens in A172, T98G, and U87MG were obtained from the
preprocessed sci-Plex 4 dataset GSM7056151 {[}7{]}. These screens
included 22 compounds at four doses, alone and in combination with
trametinib at the corresponding dose, together with trametinib-only and
vehicle controls. Each compound--dose combination was treated as a
perturbation. Preprocessing followed the K562 procedure, retaining 82,
84, and 81 pairs, respectively.

For the shared-noise analysis, we used the scPerturb release {[}24{]} of
the melanoma Perturb-CITE-seq dataset {[}20{]}. We retained cells with a
single detected guide and 217 knockouts represented by at least 40 cells
in each of the unstimulated, interferon-\ensuremath{\gamma}, and co-culture conditions.
This dataset was used to examine target construction rather than to
train or evaluate GrnPFN in the main benchmark.

For the additional drug--gene evaluation, PerturbFate GSE291147 provided
A375 cells with dual-sgRNA CRISPRi targeting 143 genes and non-targeting
controls, treated with vehicle or 1 \textmu{}M vemurafenib {[}21{]}. RNA counts
were calculated by summing labeled and unlabeled counts from the
tri-omic libraries. A cell was assigned to the guide target with the
most unique molecular identifiers when that target had at least two
identifiers and accounted for at least 75\% of the cell's guide
identifiers. For both datasets, controls were non-targeting guides with
vehicle, single conditions were targeting guides with vehicle or
non-targeting guides with drug, and pairs were targeting guides with
drug. Normalization, gene selection, and the minimum population size
followed the main benchmark.

\subsection{Construction of inputs, interaction targets, and evaluation
folds}\label{construction-of-inputs-interaction-targets-and-evaluation-folds}

Single-perturbation and control cells were randomly partitioned into
thirds to separate their contributions to inputs, training targets, and
test targets, following the general principle of sample splitting
{[}25{]}. With \(\bar{x}\) denoting mean log-normalized expression, the
first third supplied the single-perturbation input

\[\Delta_p=\bar{x}^{(input)}_p-\bar{x}^{(input)}_{\mathrm{ctrl}}.\]

The second and third supplied the corresponding training and test
interaction targets,

\[R^{(r)}_{AB}=\bar{x}_{AB}-\bar{x}^{(r)}_A-\bar{x}^{(r)}_B+\bar{x}^{(r)}_{\mathrm{ctrl}},\qquad r\in\{train,test\}.\]

The pair mean \(\bar{x}_{AB}\) used all cells of that pair. Pair
populations were disjoint between training and testing within a fold.
For inner model selection and fine-tuning, the training-target third was
further divided into halves to construct \(R_{\mathrm{fit}}\) and
\(R_{\mathrm{val}}\). The benchmark retained the 256 genes with the
largest maximum absolute single-perturbation effect in the input third.
We refer to this separation of cell roles as cross-fitted evaluation.

Each pair belonged to one of five test folds. K562 used the folds
employed for the deep-learning comparators, whereas the other screens
used a shuffled five-fold partition with random state 0. Pair folds were
held fixed across cell partitions and were not grouped by perturbation
member or compound. Across the three drug screens, 245 of 247 test pairs
had the same compound represented at another dose in the full training
set, and the matching-dose trametinib anchor was shared. In K562 and
THP-1, 271 of 273 test pairs shared at least one member with the full
training set. Single-perturbation profiles were available for all
members.

For a nominal measurement fraction \(g\) and \(n\) training pairs, the
measured count was \(b=\max(10,\mathrm{round}(gn))\). Main-benchmark
folds contained 64--114 training pairs. We evaluated
\(g\in\{0.15,0.25,0.40,1\}\), using three independent random subsets at
each partial budget and all training pairs at the full budget. GrnPFN
and fitted baselines received the same measurements in learning-curve
comparisons. Zero-shot prediction used no measured pair from the target
screen, and a 100\% budget included only the training pairs of a fold.

\subsection{Prediction scores and centering
conventions}\label{prediction-scores-and-centering-conventions}

The principal score was cosine similarity between predicted and observed
interaction vectors over the 100 genes with the largest absolute
centered observed interaction. Each method's predictions were centered
by their mean across the test pairs of the fold. Observed interactions
used one of two centers. Benchmark centering used the mean
training-target interaction of the measured pairs. Test-pair centering
used the mean observed interaction across test pairs and was applied to
zero-shot analyses and to all budgets in the learning curves.

Test labels entered only the evaluation score and the selection of
scored genes. They were not used for model fitting or tuning. Cosine
similarity measured directional agreement rather than accuracy in
interaction magnitude.

\subsection{Simulation prior}\label{simulation-prior}

We generated 140,000 screens from random gene-regulatory networks
containing 4--15 unmeasured pathway nodes, 40--99 perturbable
transcription factors, 1--6 saturating program nodes, and approximately
220--400 target genes. Regulators were chosen by preferential
attachment. Each node integrated its regulators through a logistic
response. For node \(j\) in cell \(i\),

\[y_{ij}=s_j\varepsilon_{ij}\left[\beta_j+\alpha_j\sigma\left(\sum_\ell w_{j\ell}\log(y_{i\ell}/h_\ell)+c_j\right)\right],\]

where \(y_{ij}\) was the node's activity, the sum ran over its
regulators \(\ell\), \(s_j\) was a node scale, \(\beta_j\) a basal
level, \(\alpha_j\) the maximum inducible amplitude, \(\sigma\) the
logistic function, \(w_{j\ell}\) a signed regulatory weight, \(c_j\) an
offset, and \(h_\ell\) the regulator's median in a pilot simulation of
control cells. The factor \(\varepsilon_{ij}\) combined cell-specific
log-normal intrinsic noise with sparse node loadings on cell-state
factors. A perturbed node's activity in cell \(i\) was further
multiplied by a factor depending on perturbation \(p\). CRISPR
activation used \(1+(a_p-1)u_{ip}\), with fold activation \(a_p\)
sampled log-uniformly between 3 and 50. Knockdown or inhibition used
\(1-e_pu_{ip}\), where the maximum fractional reduction \(e_p\) was
sampled uniformly between 0.4 and 0.95 for knockdown and followed a
saturating dose response for drugs. The per-cell strength \(u_{ip}\) was
a Beta-distributed value set to zero in non-responding cells, with its
distribution and the fraction of responding cells varying across
screens. In double perturbations, \(u_{ip}\) was scaled by a
shared-effector dilution factor. Measured genes were sampled as Poisson
counts with cell-specific library sizes and normalized as in the real
data.

Activation, knockdown, and drug screens were sampled with probabilities
0.4, 0.3, and 0.3. Genetic screens perturbed 20--89 transcription
factors, preferentially choosing nodes with larger out-degree, and
measured 30--150 random pairs. Drug screens applied 5--22 compounds at
2--4 doses to pathway nodes and paired each compound with a hub
inhibitor at the same dose. Population sizes followed empirical
distributions from the real screens. Single-perturbation and control
cells were assigned to the same three roles as in the benchmark, and
input single effects determined gene selection.

The hybrid prior added a screen-specific component to the
network-derived interaction:

\[R_{AB}=R^{\mathrm{GRN}}_{AB}+\rho\left[f(z_A,z_B)P+(m_A\odot m_B)Q\right].\]

Here \(z_p\) was a normalized random projection of the single profile
with rank 2--12. The function \(f\) was a random symmetric two-layer
network of \(z_A+z_B\), \(z_A\odot z_B\), and \(|z_A-z_B|\), with
\(L=2\)--8 outputs and a randomly selected activation. Independent
member effects \(m_p\sim\mathcal{N}(0,I_L)\) provided information
learnable through pairs sharing a member. The output programs \(P\) and
\(Q\) mixed single-profile directions with sparse random gene programs.
The relative component strength \(\rho\) was sampled log-uniformly over
0.3--3, and the component was absent in 25\% of screens. The same
component was added to training and test targets, preserving separate
cell-derived sampling noise in the two roles.

\subsection{Architecture and
pretraining}\label{architecture-and-pretraining}

GrnPFN processed single perturbations, measured pairs, and query pairs
on a population-by-gene grid. Tokens contained 11 features. A
single-perturbation token carried the single effect, a pair token
carried the sum, product, and absolute difference of member effects, and
a context pair carried the measured interaction. Additional features
represented the control level, token type, and log cell counts. Effect
values were divided by the screen's root-mean-square single effect.

The network comprised four layers of width 64 with four attention heads.
Each layer applied gene-axis attention within populations using 16
inducing points, population-axis attention within genes, and a
feed-forward block. Population attention read single and measured-pair
tokens, and query tokens never served as keys. Learned additive biases
distinguished unrelated populations, a pair attending to one of its
members, pairs sharing a member, and a single attending to a pair
containing it. Biases were initialized to zero. A linear head returned
the interaction prediction. The final model contained 347,521 parameters
and, without gene-identity embeddings, was equivariant to gene and
population permutations and symmetric under exchange of pair members.

Each training episode used one simulated screen. With probability 0.65,
10--30\% of pairs, up to 40, served as queries. Otherwise, 10--25\% of
perturbation members were held out, excluding the drug anchor, and all
pairs containing those members became queries. A random subset of
context pairs was retained, ranging from one third of those available to
160 pairs. Context carried training-third interactions, and the
squared-error loss used query test-third interactions. Episodes were
weighted by the inverse mean squared norm of context interactions.

Optimization used AdamW with learning rate \(3\times10^{-4}\), weight
decay 0.01, a one-cycle schedule, batch size 4, gradient clipping at
norm 1, and bfloat16 precision. The base network trained for 40,000
steps on the network prior, followed by 10,000 steps with the hybrid
prior and 10,000 further steps with membership-aware attention. A
predictive-variance head consisting of layer normalization and two
linear layers was then trained for 5,000 steps with the rest of the
model frozen. It predicted log variance per pair and gene by minimizing
Gaussian negative log-likelihood of the mean predictor's residuals.
Detached features ensured that variance training left mean predictions
unchanged. The variance included sampling variation in the targets as
well as prediction error.

\subsection{Model development and ablation
controls}\label{model-development-and-ablation-controls}

Development used cell-partition seed 301 in K562, THP-1, A172, and T98G.
U87MG was excluded from selection of the simulator, architecture, and
adaptation procedures. Each retained component was compared with a
control continued from the same checkpoint for the same number of steps
without that component. The hybrid prior was compared with continued
training on the network prior, and membership-aware attention was
compared with continued hybrid-prior training using standard softmax
attention.

The hybrid prior increased the loss of K562 accuracy after context
permutation from 0.064 to 0.156. At the full budget, its gains over the
matched control were 0.012--0.031 in K562, A172, and T98G, with a
decrease of 0.007 in THP-1. Membership-aware attention increased K562
context-permutation loss from 0.145 to 0.226 and full-budget accuracy by
0.038 (95\% CI, 0.022--0.053). Changes in the other development screens
were 0.006--0.011. Context-use experiments compared true context,
interactions permuted among context pairs, and no context while
retaining the same single profiles and pair identities.

Alternative population-attention mechanisms using an entropic
optimal-transport coupling or partner-aligned attention produced changes
of \ensuremath{-}0.007 to 0.010 and \ensuremath{-}0.008 to 0.005 across screens and budgets and
were not retained. Additional simulations with dense combination
coverage or a global-epistasis component improved some screens but
reduced accuracy in others. A member-signature prior matched several
training-data summaries more closely but reduced the inner-learning
score from 0.292 to 0.280 and full-context accuracy by a mean of 0.014.
A bilinear partner-response component concentrated drug interactions
into fewer programs than observed and was not taken forward to training.
These experiments informed model development and were separate from the
subsequent evaluation of the frozen pipeline.

\subsection{Regression and transfer
baselines}\label{regression-and-transfer-baselines}

Kernel ridge regression used a degree-2 polynomial kernel symmetrized
over pair-member order. With \(\phi_{AB}=[\Delta_A,\Delta_B]\),

\[K(AB,CD)=\frac{1}{2}\left[k_\theta(\phi_{AB},\phi_{CD})+k_\theta(\phi_{AB},\phi_{DC})\right],
\qquad k_\theta(\phi,\phi')=(\theta\phi^\top\phi'+1)^2.\]

Kernel penalties comprised seven logarithmically spaced values between
\(10^{-3}\) and \(10^3\). Seven values of \(\theta\) between \(10^{-2}\)
and 10 were scaled by the inverse mean squared input norm of measured
pairs. Kernel ridge hyperparameters were selected using four repeated
five-fold inner splits, fitting to \(R^{\mathrm{fit}}\) and validating
against \(R^{\mathrm{val}}\). Settings were ranked by the average of
their ranks for median uncentered cosine and mean pair-specific centered
cosine.

Zero-shot baselines trained on all pairs from the other four screens,
using their training-third interactions under the same cell partition.
For a pair and gene, the features were \(d_A+d_B\), \(d_Ad_B\),
\(|d_A-d_B|\), \(\min(d_A,d_B)\), \(\max(d_A,d_B)\), and
\(\log(1+\mu)\), where \(d_A\) and \(d_B\) were the members' single
effects on that gene divided by the screen's root-mean-square single
effect and \(\mu\) was the gene's control mean. Targets used the same
effect scale. Gradient boosting used scikit-learn's
HistGradientBoostingRegressor with default settings and random state 0
{[}26{]}. The linear subadditivity model predicted
\(\gamma(\Delta_A+\Delta_B)\), with \(\gamma\) fitted by pooled least
squares. For A375, transfer baselines trained on all five main screens.

\subsection{Deep-learning comparators}\label{deep-learning-comparators}

GEARS 0.1.2 {[}8{]}, CellFlow 0.0.9 {[}10{]}, and cpa-tools 0.8.1
{[}9{]} were trained with published or tutorial settings and one seed.
All three were evaluated in K562. CellFlow and CPA were evaluated in the
other main screens, and GEARS was additionally evaluated in THP-1. GEARS
was not evaluated on drug perturbations. Predicted pair effects were
converted to interactions by subtracting the input single effects and
restricting the output to the 256 benchmark genes. Models used the same
pair folds, but their training data included all single-perturbation and
control cells, including cells used to form test targets. This cell
access differed from that of GrnPFN and the fitted regression baselines.

\subsection{Confidence and perturbation
coverage}\label{confidence-and-perturbation-coverage}

For zero-shot confidence analyses, relative uncertainty was
\(\mathrm{tr}(V)/\lVert\hat{R}\rVert^2\), with predictions centered
across test pairs. Pairs were assigned to thirds within each screen and
cell partition and scored using test-pair centering. Each pair's score
was averaged across the partitions in which it entered a given third.
Intervals comparing extreme thirds resampled pairs within screens and
thirds.

For perturbation-coverage analyses, measured context was reconstructed
from each draw's seed. A genetic test pair was unseen when neither
member occurred in context. A drug test pair was unseen when its
non-anchor compound was absent at every dose. Scores were averaged per
pair within each group using benchmark centering, and method differences
were paired. Pooled estimates included screens with at least five
contributing pairs. THP-1 did not meet this threshold for unseen pairs
at 25\% and 40\%. A pair could contribute to both groups across
different measurement draws.

\subsection{Fine-tuning}\label{fine-tuning}

Fine-tuning updated a copy of GrnPFN using measured pairs only. Each
step drew two episodes, splitting measurements into 80\% pseudo-context
and 20\% pseudo-queries, with at least two query pairs. Context used
\(R^{\mathrm{fit}}\) and query losses used \(R^{\mathrm{val}}\).
Optimization used AdamW with learning rate \(10^{-4}\), no weight decay,
and gradient clipping at norm 1.

To choose the number of updates, one fifth of measured pairs was held
out. A copy tuned on the remaining pairs was assessed every 10 steps up
to 100 using pair-specific cosine. The best step count, with ties
resolved in favor of fewer steps, was then used to tune a fresh copy on
all measured pairs. Final predictions used all measured pairs with
\(R^{\mathrm{train}}\) as context. Settings were shared across screens,
and all fine-tuning experiments used the development cell partition.

Controls either permuted pseudo-context labels within each episode or
reassigned measured interactions to incorrect pairs during both tuning
and stopping. Context permutation retained 82--121\% of the gains at
15--40\% budgets in A172 and T98G, and 40\% and 91\%, respectively, at
the full budget. Incorrect pair assignments removed the benefit. A more
expensive stopping rule averaged validation curves across five inner
folds. Changes relative to the standard rule ranged from \ensuremath{-}0.018 to
0.013, and the alternative was not retained.

\subsection{Interaction reproducibility and additional-screen
eligibility}\label{interaction-reproducibility-and-additional-screen-eligibility}

Partial split-half reproducibility compared development training
interactions constructed from disjoint halves of single-perturbation and
control cells while sharing pair cells. Full split-half reproducibility
also divided pair cells. Centered cosine across 256 genes for the
partial split was 0.685, 0.157, 0.516, 0.337, and 0.299 in K562, THP-1,
A172, T98G, and U87MG, respectively. Full-split values were 0.567,
0.123, 0.387, 0.221, and 0.141. The partial split retained shared
pair-cell noise, whereas the full split reduced measurement depth, so
neither provided an independent experimental noise ceiling.

In the melanoma noise analysis, the unstimulated knockout profile was
calculated either from the cells contributing to the interaction target
or from a disjoint half. The vanishing-effect predictor returned the
negative of that input, and evaluation used the 100 most changed genes.

Before evaluating additional drug--gene datasets, we required a mean
training-pair split-half cosine above 0.10 using independent halves of
all four cell populations and the 256 selected genes. A375 evaluation
used the frozen model, five pair folds, and the same four cell-partition
seeds and prediction settings as the main benchmark. The dataset was not
used for model development.

\subsection{Statistical analysis}\label{statistical-analysis}

The zero-shot comparison against gradient boosting was specified before
evaluation on the four new partitions. Its 97.5\% confidence interval
reflected Bonferroni adjustment for two endpoints. The 15\% comparison
against kernel ridge was specified separately, after the GrnPFN scores
from the first evaluation had been inspected but before fitted baselines
were run, and used a 95\% interval. These analyses were therefore not
two co-primary endpoints of a single blinded test. Confidence,
perturbation-coverage, additional-screen, and dependence-aware analyses
were specified before their own execution but after earlier results were
known and were treated as descriptive.

For inference, scores were averaged per test pair across applicable
measurement draws and cell partitions. Per-screen method differences
used 2,000 paired bootstrap resamples of pairs. Pooled estimates gave
each screen equal weight and resampled pairs within screens.

Approximate dependence-aware intervals resampled compounds in drug
screens while retaining their doses together. In genetic screens,
perturbation members received multinomial multiplicities, and each pair
received the product of its two members' weights, following
product-weight bootstrap principles {[}27{]}. Pooled estimates again
used equal screen weights and 2,000 resamples. This procedure addressed
dependence from shared members or compounds but did not account for
every shared control or anchor effect. The four cell partitions reused
the same biological material and were not treated as independent
biological replicates.

All eligible pairs were retained without a prospective sample-size
calculation. Cell partitions and measurement subsets used seeded
generators, and scores were calculated programmatically.

\section{Data and code availability}\label{data-and-code-availability}

Source datasets are available from Gene Expression Omnibus under
GSE133344, GSE213957, GSM7056151 (within GSE225775), GSE291147, and
GSE280506. The melanoma Perturb-CITE-seq dataset is available through
scPerturb and Broad Single Cell Portal study SCP1064. The project
repository is \url{https://github.com/haihuilab/GrnPFN}.

\section{References}\label{references}

\protect\phantomsection\label{refs}

\end{document}